%% file: manuscript.tex
\documentclass[sigconf, authorversion,nonacm,screen]{acmart}
\AtBeginDocument{%
  }

\setcopyright{none}
\acmConference[ICSE '26]{}{April 12--18, 2026}{Rio De Janeiro, Brazil}
\usepackage{microtype}
\usepackage{booktabs} % for formal tables
\usepackage{adjustbox}
\usepackage[frozencache,cachedir=minted-cache]{minted}
\usepackage{array}
\usepackage{paralist} % compactitem lists
\usepackage{graphicx}
\usepackage{mdframed} % for colored frames, for example for pretty-styled RQs
\usepackage{enumitem}
\usepackage{multirow}
\usepackage{makecell}
\usepackage{enumitem}

\newcommand\RQAnswer[2]{
	\noindent 
	\begin{mdframed}[
		linewidth=1pt,
		leftmargin=0pt,
		backgroundcolor=gray!10,
		linecolor=gray!10]
		\begin{tabular}{@{}l m{0.89\textwidth}}
			\textbf{{#1}} & {#2} \\
		\end{tabular}
	\end{mdframed}
}

\definecolor{clean}{HTML}{1F77B4}
\definecolor{confusing}{HTML}{D62728}

\newcommand\clean{\textcolor{clean}{clean~}}
\newcommand\confusing{\textcolor{confusing}{confusing~}}
\newcommand\Clean{\textcolor{clean}{Clean~}}
\newcommand\Confusing{\textcolor{confusing}{Confusing~}}

\begin{document}

% \title{Beyond Atoms of Confusion: A Perplexity-Based Approach to Identifying Confusing Code Regions}

\title[How do Humans and LLMs Process Confusing Code?]{How do Humans and LLMs Process Confusing Code?}

\author{Youssef Abdelsalam}
\email{abdelsalam@cs.uni-saarland.de}
\orcid{0009-0009-4444-765X}
\affiliation{%
  \institution{Saarland University \\ Saarland Informatics Campus}
  \city{Saarbrücken}
  %\state{Saarland}
  \country{Germany}
}

\author{Norman Peitek}
\email{peitek@cs.uni-saarland.de}
\orcid{0000-0001-7828-4558}
\affiliation{%
  \institution{Saarland University \\ Saarland Informatics Campus}
  \city{Saarbrücken}
  %\state{Saarland}
  \country{Germany}
}

\author{Anna-Maria Maurer}
\email{maureran@cs.uni-saarland.de}
\orcid{0009-0002-0379-5534}
\affiliation{%
  \institution{Saarland University \\ Saarland Informatics Campus}
  \city{Saarbrücken}
  %\state{Saarland}
  \country{Germany}
}

\author{Mariya Toneva}
\email{mtoneva@mpi-sws.org}
\orcid{0000-0002-2407-9871}
\affiliation{%
  \institution{Max Planck Institute for \\ Software Systems}
  \city{Saarbrücken}
  %\state{Saarland}
  \country{Germany}
}

\author{Sven Apel}
\email{apel@cs.uni-saarland.de}
\orcid{0000-0003-3687-2233}
\affiliation{%
  \institution{Saarland University \\ Saarland Informatics Campus}
  \city{Saarbrücken}
  %\state{Saarland}
  \country{Germany}
}

\renewcommand{\shortauthors}{Abdelsalam et al.}

\begin{abstract}
% background
Already today, humans and programming assistants based on large language models (LLMs) collaborate in everyday programming tasks.
Clearly, a misalignment between how LLMs and programmers comprehend code can lead to misunderstandings, inefficiencies, low code quality, and bugs. 
% research question
A key question in this space is whether humans and LLMs are confused by the same kind of code. This would not only guide our choices of integrating LLMs in software engineering workflows, but also inform about possible improvements of LLMs.
% method
To this end, we conducted an empirical study comparing an LLM to human programmers comprehending clean and confusing code. We operationalized comprehension for the LLM by using \emph{LLM perplexity}, and for human programmers using \emph{neurophysiological responses} (in particular, EEG-based fixation-related potentials). 
% results
We found that LLM perplexity spikes correlate both in terms of location and amplitude with human neurophysiological responses that indicate confusion. This result suggests that LLMs and humans are similarly confused about the code. Based on these findings, we devised a data-driven, LLM-based approach to identify regions of confusion in code that elicit confusion in human programmers.
\end{abstract}

%%
%% The code below is generated by the tool at http://dl.acm.org/ccs.cfm.
%% Please copy and paste the code instead of the example below.
%%


%%
%% Keywords. The author(s) should pick words that accurately describe
%% the work being presented. Separate the keywords with commas.

%\received{11 July 2025}
%\received[revised]{12 March 2009}
%\received[accepted]{5 June 2009}

\maketitle

\input{introduction}
\input{background_and_relatedwork}
\input{research_questions}
\input{methodology}

\input{results}
\input{discussion}

\input{threats}
\input{conclusion}

\input{acknowledgments.tex}

%%
%% The next two lines define the bibliography style to be used, and
%% the bibliography file.
\bibliographystyle{ACM-Reference-Format}
\bibliography{bibliography}

%%
%% If your work has an appendix, this is the place to put it.
\appendix

\end{document}

%% file: introduction.tex
\section{Introduction}
\label{sec:introduction}

Large language models (LLMs) have rapidly advanced and demonstrated remarkable performance in many programming-related tasks~\cite{hou2024large}, including code completion~\cite{husein2025large, izadi2022codefill}, synthesis~\cite{nguyen2022copilot}, and repair~\cite{hossain2024deepdive,zubair2025llmrepair}. 
Trained on massive corpora of data, LLMs are able to generate syntactically correct and semantically plausible code across all areas of software engineering, from education~\cite{raihan2025LLMeducation, rasnayaka2024LLMProject, Hellas2023, Kazemitabaar2024} and research~\cite{demartino2024frameworkusingllmsrepository,felizardo2024literatureReview,steinmacher2024EmulateHuman} to practice~\cite{khojah2024codegenerationobservationalstudy,hao2024sharedconversations}. Already today, LLM-based coding assistants and human programmers work in concert to produce substantial amounts of source code~\cite{daniotti2025usingaicodeglobal}.
Clearly, a misalignment between LLM and human programmers in such a collaborative setting can lead to misunderstandings, inefficiencies,
low code quality, and bugs~\cite{ziegler2022ProductivityCopilot,Vaithilingam2022usabilitycodegen, Siddiq2024}. To maximize the potential of LLMs, we are well advised to critically assess the alignment~\cite{moussa2023improving, alkhamissi2025llmsoutgrow, sergeyuk2024reassessing, barke2023} of how LLMs and humans comprehend source code.

\begin{figure}[t]
\centering
    \includegraphics[width=\columnwidth]{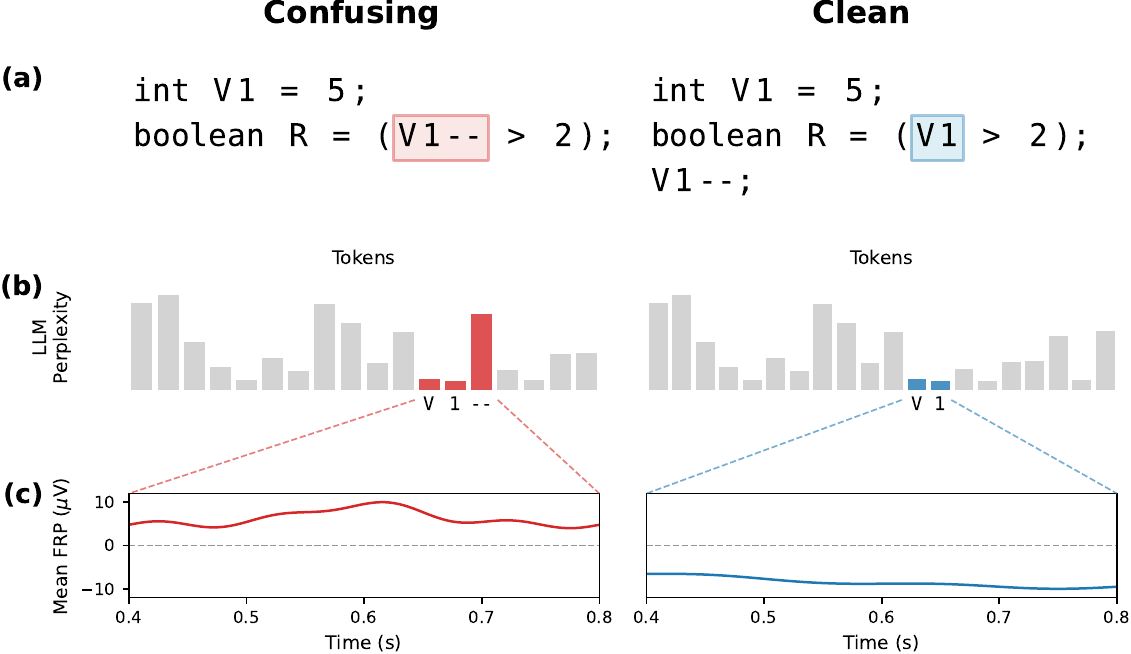}
    \caption{A comparison of \confusing and corresponding \clean code (a) as well as effects on LLM perplexity (b) and neurophysiological responses (c):
    The \confusing code snippet includes an atom of confusion (highlighted in red). The corresponding \clean snippet avoids this atom by extracting the decrement-expression into a separate statement (a). The token-level perplexity from an LLM increases for the confusing code region (b). Corresponding EEG fixation-related potentials exhibit increased amplitude during the comprehension of the confusing snippet (c). 
    }
    \label{fig:intro_figure}
\end{figure}

The premise of our investigation is that understanding (the presence and extent of) this alignment would allow us to better understand possible risks and pitfalls of integrating LLMs into software engineering workflows and to improve the capabilities of LLMs in collaborating with humans in programming-related tasks. 
Specifically, we are interested in the reaction of LLMs and humans to \emph{confusing} code. Prior research has dedicated considerable attention to problematic or confusing code~\cite{Dijkstra1968GoTo, gopstein2017atoms}, even offering validated stimuli that elicit distinct neurophysiological responses in humans~\cite{yeh2017eeg, yeh2022eeg, bergum2025unexpectedinformative}. Thus, confusing code provides an empirically grounded and well-studied setting to evaluate the alignment between human and LLM comprehension.

\paragraph{Research questions and methodology.}
Our investigation is guided by three research questions:

\begin{itemize}[leftmargin=5.9ex]
    \item[RQ\textsubscript{1}:] How does confusing code affect LLM confusion?
    \item[RQ\textsubscript{2}:] How does LLM confusion about code relate to human confusion about the same code?
    \item[RQ\textsubscript{3}:] How well is a data-driven, LLM-based approach able to identify code regions that confuse humans?
\end{itemize}

For our investigation, we rely on a well-studied set of syntactically valid and semantically correct code idioms that are known to confuse human programmers, called \emph{atoms of confusion}~\cite{gopstein2017atoms}. \autoref{fig:intro_figure}(a) displays an example pair of an atom of confusion in a confusing and a corresponding clean (i.e., less confusing) code snippet. Atoms of confusion lead to longer reading times, higher error rates, and lower comprehension accuracy across programming languages~\cite{langhout2021atoms, daCosta2023confusion, gopstein2017atoms}. Studies involving visual attention~\cite{olivera2020eye, daCosta2023confusion} and neurophysiological measures using electroencephalography (EEG)~\cite{yeh2017eeg, yeh2022eeg} further corroborate these results. 

To analyze how humans comprehend code, we rely on EEG data from prior work. EEG has been a well-established method for measuring programmers' neurophysiological responses to source code~\cite{yeh2017eeg,peitek2022efficacyEEG,ishida2019EEG,lee2016noviceEEG, Crk2014, Medeiros2021, Lin2021}. In a study on atoms of confusion, \citeauthor{bergum2025unexpectedinformative} combined EEG with eye tracking to measure fixation-related potentials (FRPs). They found that confusing code reliably elicits a so-called \emph{late frontal positivity} in the EEG signal of human programmers (see \autoref{fig:intro_figure}(c)), which is a neurophysiological correlate of human confusion~\cite{kuperberg2020twoposn400}. Therefore, we also measure human confusion using fixation-related potentials with the focus on the amplitude of late frontal positivity.

To analyze how LLMs comprehend code, we use the measure of \emph{perplexity}. Perplexity reflects how unexpected a token is to the LLM given its context (see \autoref{fig:intro_figure}(b)). In natural language processing and psycholinguistics, the unexpectedness of natural language text serves as a well-established proxy for human confusion and cognitive load~\cite{LEVY20081126}, which has been confirmed also for source code~\cite{hindle2012naturalness, casalnuovo2020difficulty}.

Combining the pieces, we have devised a methodology that: (1) computes token- and snippet-level perplexity values with state-of-the-art LLMs, (2) extends this analysis to calculate perplexity at predefined areas of interest (AOIs), (3) aligns these with fixation-related potentials in EEG signals, and (4) detects perplexity peaks within code snippets to define new AOIs, extract corresponding FRPs, and align these measures accordingly. We leverage the dataset of \citeauthor{bergum2025unexpectedinformative}~\cite{bergum2025unexpectedinformative} for fixation-related potentials and the state-of-the-art LLM \texttt{Qwen2.5-Coder-32B}~\cite{hui2024qwen2} for perplexity measurement.

\paragraph{Results.}
Our results demonstrate that the presence of atoms of confusion consistently triggers a spike in LLM perplexity, revealing that LLMs are also sensitive to these confusing code constructs, corroborating prior findings on human programmers. Moreover, when correlating the LLM perplexity values with the strength of human confusion measured by the amplitude of corresponding fixation-related potentials, we find a consistent, statistically significant positive relationship. The confusion of LLMs and humans is not only spatially aligned (based on tokens), but also in terms of their amplitude. 
Based on these findings, we have devised a data-driven, LLM-based approach to identify code regions of confusion that are associated with human confusion.
%Further, we provide evidence that LLM perplexity can serve as a scalable proxy for identifying confusing code regions, bridging computational and neurophysiological perspectives on confusion in code.

Overall, we find an alignment between LLM perplexity and human neurophysiological responses, which is an important step toward establishing LLMs as viable surrogate models of human program comprehension. This paves the way to deepen our understanding of how programmers comprehend code and supports the development of LLMs to be more closely aligned with human comprehension patterns. Our investigations suggest that LLMs may serve as a foundation for data-driven feedback methods that flag or revise cognitively demanding code based on LLM perplexity and facilitate the development of tools that automatically refactor confusing code or highlight areas where programmers may need additional information to comprehend the code.

\paragraph{Contributions.}
We make the following contributions:

\begin{itemize}[leftmargin=3ex]
    \item An investigation of the effects of confusing code (based on atoms of confusion) on LLM perplexity;
    \item A novel combined analysis of LLM perplexity and human confusion measured in terms of fixation-related potentials, showing a positive correlation between LLM perplexity and neurophysiological responses indicative of human confusion;
    \item A data-driven approach to detect potential regions of confusion using LLM perplexity, validated with EEG data to confirm the effect of the code region on human confusion; and
    \item A replication package with our data, the entire analysis pipeline, and additional details.\footnote{\url{https://github.com/brains-on-code/llm-perplexity-AoC}}
\end{itemize}

%% file: background_and_relatedwork.tex
\section{Background and Related Work}
\label{sec:background_related_work}

% Atoms of Confusion (AoCs):
% - Definition, examples, and prior findings on comprehension impact.
% - Behavioral and neural evidence of increased cognitive load (EEG, eye tracking).

% Neural Markers of Confusion:
% - EEG components linked to confusion (P600 late frontal positivity).
% - Combined EEG and eye tracking to localize confusion markers (FRPs).

% Perplexity as a measure of confusion:
% - perplexity as a measure of model uncertainty.
% - Prior work using perplexity in code analysis and comprehension studies.

% Research Gap: (Positioning this work)
% - Fills gap by quantitatively linking LLM perplexity with EEG FRPs on AoCs.
% - Proposes data-driven RoC detection covering (or even improving / extending upon) predefined AoCs.

% TODO: Introduction Sentence + Overview 
In the following, we provide the background and summarize prior work related to our study. We review three key concepts: (1) atoms of confusion, a set of small code patterns known to impair human comprehension despite being syntactically and semantically valid; (2) neurophysiological measures, particularly EEG and fixation-related potentials, which offer fine-grained insights into the neurophysiological responses underlying program comprehension and confusion; and (3) LLM perplexity, a measure of model uncertainty that has been linked to human difficulty in processing both natural language and source code. Together, these lines of work provide the theoretical and empirical grounding for our investigation into how LLMs and human programmers respond to confusing code.

\subsection{Atoms of Confusion}

Over the past decade, atoms of confusion (AoCs) have emerged as a key concept for understanding the cognitive challenges programmers face when reading source code. These atoms are short, syntactically and semantically valid code snippets that often tend to cause confusion. First introduced by \citeauthor{gopstein2017atoms}, atoms of confusion were identified through behavioral studies in C and C++ that revealed certain patterns, such as misleading variable shadowing or misleading increment placement (see Figure~\ref{fig:intro_figure}a), that frequently led to human confusion during program comprehension tasks~\cite{gopstein2017atoms}.
Subsequent work extended the study of atoms of confusion to other programming languages. \citeauthor{langhout2021atoms} adapted the original atoms to Java and found that they led to significantly higher error rates and increased perceived difficulty~\cite{langhout2021atoms}. Da Costa et al. identified similar effects in Python, showing that atoms of confusion increase the visual effort and comprehension difficulty~\cite{daCosta2023confusion}.

\renewcommand{\arraystretch}{1.1}  % slightly tighter row height

\begin{table}[h]
\centering
\caption{Illustrative \confusing code (atoms of confusion) and \clean counterparts based on \citeauthor{bergum2025unexpectedinformative}~\cite{bergum2025unexpectedinformative}}
\label{tab:atoms_of_confusion}
\begin{adjustbox}{width=\columnwidth}
\footnotesize
\begin{tabular}{@{}l l l @{}}
\toprule
\textbf{Atom of Confusion} & \textbf{\Confusing Variant} & \textbf{\Clean Variant} \\
\midrule
Arithmetic as Logic & 
\begin{minipage}[t]{2.5cm}
\mintinline[fontsize=\footnotesize]{java}|(V1 + 5 != 0);|
\end{minipage} & 
\begin{minipage}[t]{2.5cm}
\mintinline[fontsize=\footnotesize]{java}|(V1 != -5);|
\end{minipage} \\[0.5ex]

Change of Literal Encoding & 
\begin{minipage}[t]{2.5cm}
\mintinline[fontsize=\footnotesize]{java}|12 & 3;|
\end{minipage} & 
\begin{minipage}[t]{2.5cm}
\mintinline[fontsize=\footnotesize]{java}|0b1100 & 0b0011;|
\end{minipage} \\[0.5ex]

Constant Variables & 
\begin{minipage}[t]{2.5cm}
\mintinline[fontsize=\footnotesize]{java}|V1 = V2;|
\end{minipage} & 
\begin{minipage}[t]{2.5cm}
\mintinline[fontsize=\footnotesize]{java}|V1 = 5;|
\end{minipage} \\[0.5ex]

Dead, Unreachable, Repeated & 
\begin{minipage}[t]{2.5cm}
\mintinline[fontsize=\footnotesize]{java}|R = 3; R = 2;|
\end{minipage} & 
\begin{minipage}[t]{2.5cm}
\mintinline[fontsize=\footnotesize]{java}|R = 2;|
\end{minipage} \\[0.5ex]

Implicit Predicate & 
\begin{minipage}[t]{2.5cm}
\mintinline[fontsize=\footnotesize]{java}|!(V1 - V2 < 6)|
\end{minipage} & 
\begin{minipage}[t]{2.5cm}
\mintinline[fontsize=\footnotesize]{java}|(V1 - V2 > 5)|
\end{minipage} \\[0.6ex]

Remove Indentation & 
\begin{minipage}[t]{2.5cm}
\begin{minted}[fontsize=\footnotesize]{java}
while (V1 > 0)
   V1--;
   R++;
\end{minted}
\end{minipage} & 
\begin{minipage}[t]{2.5cm}
\begin{minted}[fontsize=\footnotesize]{java}
while (V1 > 0)
   V1--;
R++;
\end{minted}
\vspace{0.5ex}
\end{minipage} \\[0.5ex]

Infix Operator Precedence & 
\begin{minipage}[t]{2.5cm}
\mintinline[fontsize=\footnotesize]{java}|2 - 4 / 2;|
\end{minipage} & 
\begin{minipage}[t]{2.5cm}
\mintinline[fontsize=\footnotesize]{java}|2 - (4 / 2);|
\end{minipage} \\[0.5ex]

Omitted Curly Braces & 
\begin{minipage}[t]{2.5cm}
\mintinline[fontsize=\footnotesize]{java}|for (_) R++; R++;|
\end{minipage} & 
\begin{minipage}[t]{2.5cm}
\mintinline[fontsize=\footnotesize]{java}|for (_) { R++; } R++;|
\end{minipage} \\[0.5ex]

Post-Increment/Decrement & 
\begin{minipage}[t]{2.5cm}
\mintinline[fontsize=\footnotesize]{java}|int R = 3 + V1++;|
\end{minipage} & 
\begin{minipage}[t]{2.5cm}
\mintinline[fontsize=\footnotesize]{java}|int R = 3 + V1; V1++;|
\end{minipage} \\[0.5ex]

Pre-Increment/Decrement & 
\begin{minipage}[t]{2.5cm}
\mintinline[fontsize=\footnotesize]{java}|int R = ++V1 - 2;|
\end{minipage} & 
\begin{minipage}[t]{2.5cm}
\mintinline[fontsize=\footnotesize]{java}|++V1; int R = V1 - 2;|
\end{minipage} \\[0.5ex]

Type Conversion & 
\begin{minipage}[t]{2.5cm}
\mintinline[fontsize=\footnotesize]{java}|(byte) V1;|
\end{minipage} & 
\begin{minipage}[t]{2.5cm}
\mintinline[fontsize=\footnotesize]{java}|(byte) (V1 % 256);|
\end{minipage} \\
\bottomrule
\end{tabular}
\end{adjustbox}
\end{table}

Studies using mining software repository approaches have shown that atoms of confusion are not only prevalent in practice in large-scale projects, such as the Linux kernel, but are also frequently discussed or removed in later commits~\cite{gopstein2018mining}. Moreover, mining-based analyses of Java repositories further linked atoms of confusion to bug-prone code and subsequent maintenance activity, highlighting their real-world relevance~\cite{mendes2022javarepomining, pinheiro2023relateandleave}.

Despite these consistent findings, atoms of confusion remain under-represented in programming guidelines~\cite{mendes2022javarepomining}. A programmer survey has shown that, while a majority of programmers recognize certain patterns as confusing, they are often reluctant to refactor code that is syntactically correct and functionally complete~\cite{medeiros2019survey}. A follow-up study using a think-aloud protocol suggests that behavioral accuracy may not always indicate true understanding: Some participants use external cues---such as test outputs---to guess correct results even when their mental models are incorrect~\cite{gopstein2023thinkaloud}.

\subsection{Neurophysiological Correlates of Confusion}

Detailed insights into program comprehension can be gained by analyzing the behavior of programmers on a neurophysiological level.
\emph{Electroencephalography} (EEG) is a non-invasive method for capturing neural activity with high temporal resolution, making it particularly well-suited for investigating the time course of neurophysiological responses when humans are interpreting stimuli~\cite{jackson2014eeg}.
To reduce noise, it is established to average EEG responses from multiple trials time-locked to the start events, called \emph{event-related potentials} (ERPs)~\cite{luck2014erp}.
For simple stimuli, ERPs are created time-locked onto stimulus onset, thus capturing the initial neurophysiological responses of participants regarding the stimuli.
In contrast, for more complex stimuli, the setup can be extended with an eye tracker synchronized to the EEG, which allows the capture of neurophysiological responses relative to fixations in a region of interest, thereby calculating \emph{fixation-related potentials (FRPs)}~\cite{hutzler2007welcome}.
For both ERP and FRP, the EEG signal must be leveled to a baseline prior to aggregation, and commonly, the resulting ERP and FRP waves are compared in a contrast between multiple conditions in form of difference waves~\cite{luck2014erp}.
These difference waves capture the difference in neurophysiological activity at specific time points, and, to an extent, also scalp locations. This allows to observe which ERP components (i.e., established cognitive processes measured in the brain) are involved or show increased activity during the comprehension process.
Multiple ERP components have been identified in prior research across many fields~\cite{donogue2022metaerp}, including ones that specifically occur when processing natural language~\cite{swaab2012language}.
For example, one ERP component is the N400, a negative impact at about 400\,ms after the event, which reacts to semantic incongruency~\cite{kuperberg2016separate, kuperberg2020twoposn400}.

Recently, researchers have started to adapt these methods to software engineering settings. They have measured program comprehension using ERPs and FRPs and mapped the identified components to those during comprehending natural language~\cite{bergum2025unexpectedinformative, kuo2024violationserp}. 
\citeauthor{kuo2024violationserp} combined an ERP approach with token-wise presentation of code snippets to detect that programmers react to syntax and semantic violations in code with the same ERP components as when reading text in natural language~\cite{kuo2024violationserp}. \citeauthor{bergum2025unexpectedinformative} analyzed the neurophysiological response of programmers to atoms of confusion while determining the output of a code snippet using an FRP approach. They identified that these responses correspond to \emph{late frontal positivity}~\cite{bergum2025unexpectedinformative}. This is a positive ERP component concentrated on the frontal part of the scalp and peaks at about 600\,ms. It has been linked to unexpected but plausible sentence endings in natural language. They commonly occur in sentences that generate high expectations (i.e., high probability) for one specific sentence ending, but where the actual ending does not match that expectation (i.e., low probability)~\cite{kuperberg2020twoposn400,Thornhill2012, brothers2020going,van2012prediction}. For example, in the sentence \textit{he bought her a pearl necklace for her ...}, people commonly expect \textit{collection}. When a less expected ending, such as \textit{birthday}, is presented, the neurophysiological response exhibits an increased amplitude of the late frontal positivity ERP component. 
Overall, these initial studies show fundamental similarities in the neurophysiological responses of natural language and source code.

\subsection{LLM Perplexity as a Measure of Confusion}
\label{sec:perplexity_background}

Perplexity is an established metric for evaluating language models by quantifying their uncertainty in predicting sequences of tokens based on their probability~\cite{xiong2024efficient,shorinwa2025surveyuncertainty}. Concretely, a language model defines probability distributions over sequences, enabling it both to generate plausible sentences and to evaluate the likelihood of existing ones~\cite{jurafsky2025}. A well-trained language model should assign higher probabilities to coherent and well-formed sequences, and lower probabilities to unexpected ones. Formally, given a token sequence \( T = (t_1, t_2, \ldots, t_n) \), the language model assigns a joint probability:
\[
\text{Pr}(T) = \prod_{i=1}^n \text{Pr}(t_i \mid t_1, \ldots, t_{i-1}).
\]

Because this product decreases exponentially with sequence length regardless of sequence quality, a direct comparison of sequence probabilities across different token/sequence lengths is not meaningful~\cite{jurafsky2025}. To address this issue, the \emph{geometric mean} of the token probabilities is used to normalize the sequence probability
\[
\text{Pr}_{\mathrm{norm}}(T) = \text{Pr}(T)^{1/n},
\]

\noindent which corresponds to the average per-token likelihood. \emph{Average Perplexity} is then defined as the inverse of this probability:
\[
\text{PPL}(T) = \frac{1}{\text{Pr}_{\mathrm{norm}}(T)} = \left(\frac{1}{\text{Pr}(T)}\right)^{1/n}.
\]

This formulation highlights that perplexity quantifies the average uncertainty of the language model when predicting each token in the sequence, independent of sequence length~\cite{casalnuovo2020difficulty}. 

In natural language, lower perplexity indicates more predictable constructs, whereas higher perplexity reflects greater uncertainty and difficulty in prediction~\cite{xu2024investigatingefficacyperplexitydetecting, SMITH2013302, LEVY20081126}. These effects are well-established in psycholinguistics, where higher perplexity has been linked to slower reading times and increased cognitive load~\cite{SMITH2013302, LEVY20081126}.

In the context of source code, perplexity similarly reflects how ``natural'' or predictable the code is to the model~\cite{hindle2012naturalness}.
Code exhibits strong statistical regularities, much like natural language, but with even greater predictability, as programmers tend to favor idiomatic and repetitive patterns~\cite{casalnuovo2020difficulty}.
\citeauthor{casalnuovo2020difficulty} experimentally tested whether higher perplexity in code actually relates to difficulty in humans understanding code. In a controlled user study, programmers were shown functionally identical code snippets with different perplexities. They found that the versions with higher perplexity led to slower response times and lower accuracy on comprehension questions~\cite{casalnuovo2020difficulty}. \citeauthor{goodkind-bicknell-2018-predictive} even showed that a token's perplexity linearly predicts human reading time~\cite{goodkind-bicknell-2018-predictive}. 

Building on these insights, recent research has further employed token-level perplexity to assess the predictability of individual tokens rather than entire code snippets and enable a fine-grained analysis~\cite{xu2024investigatingefficacyperplexitydetecting, Xu_Sheng_2024}. For instance, \citeauthor{xu2024investigatingefficacyperplexitydetecting} used this approach to distinguish between human-written and LLM-generated code, observing that code produced by LLMs exhibits more uniform perplexity~\cite{xu2024investigatingefficacyperplexitydetecting}.

These findings suggest that perplexity not only captures structural regularities in code, but also reacts to unexpected or complex tokens on a fine-grained level, making it a practical and cognitively relevant metric for code analysis. However, the use of token-level perplexity to directly model human neurophysiological responses during program comprehension remains largely unexplored.

% add transition sentence to make focus on token-level perplexity and not on actual research
%One thread of research uses perplexity to detect unusual or AI-generated code. For example, \citeauthor{xu2024investigatingefficacyperplexitydetecting} use an LLM’s log-probabilities to compute token-level perplexity and attempt to distinguish human-written code from code generated by another LLM~\cite{xu2024investigatingefficacyperplexitydetecting}. They note that code written by language models tends to be more uniform (lower perplexity) than human code, and initial results suggest perplexity-based detectors generalize well across languages and model families~\cite{xu2024investigatingefficacyperplexitydetecting}. \citeauthor{Xu_Sheng_2024} report that subtle semantics-preserving changes cause ``higher perplexity to the model, making the code appear more unnatural''~\cite{Xu_Sheng_2024}, again indicating that language models can flag atypical code patterns via spikes in perplexity. However, these studies also note that perplexity is not a silver bullet. The same analysis found that, while perplexity detectors can generalize (e.g., across languages), their raw accuracy and efficiency is sometimes low or may be inconsistent~\cite{Xu_Sheng_2024}. Another recent study on data curation for code models observed that simply selecting code examples by perplexity (as a ``complexity'' measure) performed about as well as random selection~\cite{wang2024codellmsperformempowering}, suggesting that perplexity alone may not capture all dimensions of code difficulty for model training.

\subsection{Research Gap}

Prior work has established that atoms of confusion negatively impact program comprehension across several programming languages. They occur frequently in software repositories and are associated with increased error rates, maintenance effort, and bug-prone code.
It has been demonstrated that LLMs assign higher perplexity to certain code patterns, and that this perplexity correlates in cases with human difficulty during comprehension tasks.
Recent studies using neurophysiological measures such as fixation-related potentials have shown that confusing code elicits distinct neurophysiological responses, particularly late frontal positivity, indicating increased mental effort during comprehension. Both LLM perplexity and late frontal positivity have one fundamental aspect in common: They are affected by what LLMs, respectively humans, perceive as a probable sequence based on the previously given context and training or experience.

Together, these lines of work suggest that there may be meaningful overlap between patterns that confuse humans, surprise language models, and trigger distinct neurophysiological responses. While atoms of confusion are known to induce confusion, and perplexity has been linked to comprehension difficulty in isolated studies, it is unclear whether LLMs assign higher perplexity to these known confusing constructs or whether LLM perplexity aligns with human neurophysiological responses during program comprehension. If such alignment exists, it raises the question of whether we can move beyond static, predefined patterns and use perplexity itself to automatically identify \emph{regions of confusion} in code.

% Demonstrating this alignment is an important step toward establishing LLMs as viable cognitive models of program comprehension. More broadly, aligning LLM perplexity with human neurophysiological responses can deepen our understanding of how programmers process code and guide the development of LLMs that better reflect human comprehension patterns. This alignment lays the foundation for novel feedback mechanisms that identify cognitively demanding code segments through LLM perplexity, thereby supporting tools that automatically refactor confusing code or highlight areas where programmers may require additional assistance.

%% file: research_questions.tex
\section{Research Questions}
\label{sec:research_questions}

Based on this research gap, we structure our investigation around three main research questions, each exploring a different aspect of the relationship between confusing code constructs, LLM confusion, and human neurophysiological responses observed during program comprehension.

\paragraph{RQ\textsubscript{1}: How does confusing code affect LLM confusion?}

RQ\textsubscript{1} addresses whether LLM confusion reflects known cognitive challenges posed by confusing code. Here, we operationalize confusing code using the established atoms of confusion and LLM confusion to this code with perplexity. To establish perplexity as a computational measure of confusion, we compare the perplexity of code snippets containing atoms of confusion to their clean, functionally equivalent counterparts.

Based on the literature (cf.~Section~\ref{sec:background_related_work}), we expect code snippets containing atoms of confusion to exhibit (1) higher perplexity particularly for the regions containing the atoms of confusion (as shown in \autoref{fig:intro_figure}(b)), and (2) statistically significant differences in perplexity metrics (i.e., average and maximum) between confusing and clean versions of the code, especially within the regions that contain the atoms of confusion.

\paragraph{RQ\textsubscript{2}: How does LLM confusion about code relate to human confusion about the same code?}

Building on RQ\textsubscript{1}, we investigate the extent to which LLM perplexity aligns with neurophysiological correlates of human confusion during program comprehension. To this end, we analyze the EEG recordings and eye-tracking data of \citeauthor{bergum2025unexpectedinformative}~\cite{bergum2025unexpectedinformative}, in which they identified a late frontal positivity using fixation-related potentials. This neurophysiological response, often associated with increased syntactic complexity and processing effort~\cite{kuperberg2020twoposn400}, provides a robust basis for evaluating whether code regions with high LLM perplexity correspond to increased cognitive demand in human programmers.

Specifically, we analyze whether the maximum token-level perplexity in atoms of confusion correlates with increased late frontal positivity in the FRP signal that is extracted time-locked to fixations positioned on this token sequence. We hypothesize that the LLM perplexity will positively correlate with the measured late frontal positivity, suggesting that LLM perplexity and human confusion are linked.

\paragraph{RQ\textsubscript{3}: How well is a data-driven, LLM-based approach able to identify code regions that confuse humans?}

Finally, we explore whether perplexity can serve as a practical proxy for identifying confusing code regions by testing (1) whether regions of confusion derived from perplexity align with neurophysiological correlates of human confusion as measured by late frontal positivity, and (2) whether these detected regions of confusion result in similar correlations with EEG signals compared to known atoms of confusion.

Our third research question is concerned with the feasibility of using LLM perplexity directly to identify confusing regions in code, beyond the predefined set of atoms of confusion. For this purpose, we introduce a data-driven approach for extracting \emph{regions of confusion} by locating peaks in token-level perplexity and aggregating them into semantically coherent sequences of tokens. These regions are then compared to the predefined atoms of confusion in terms of their alignment with FRP responses.
We hypothesize that regions of confusion extracted through perplexity-based methods will yield correlations that are, at least, comparable, and possibly stronger, than those based on the predefined set of atoms of confusion---offering a scalable way to localize confusion in code.

%% file: methodology.tex
\section{Methodology}
\label{sec:methodology}

% Data:
% - 72 Java code snippets annotated with predefined AoCs, clean and confusing versions.
% - EEG Data: data collection, preprocessing (downsampling, epoch extraction).
% - Eye-Tracking Data: data collection, preprocessing.
% - Combining EEG and eye tracking (setting FRP markers).

% Perplexity Analysis:
% - LLM per-token perplexity calculation (detailed tokenization, AOI parsing, etc.).
% - Model comparison and choice of model to be used.
% - Metrics: overall, mean, max perplexity, AOI-specific perplexity.
% - Statistical tests comparing clean vs confusing code perplexity and EEG.
% - Correlation analyses between perplexity and EEG signal.

% Data-driven RoC detection:
% 1. Peak Detection: Identify peaks in perplexity values using a peak detection algorithm (based on prominence, distance, and span).
% 2. Lexical and Syntactic Expansion: Expand the detected regions by combining related tokens (e.g., variable names and numbers).
% 3. Region Merging: Merge adjacent or overlapping regions.

In this section, we present our research process, including the extraction and processing of the EEG data, the selection of language models, the perplexity analysis, and the approach to generate regions of confusion based on a data-driven approach.

\subsection{EEG Data on Human Confusion}

The dataset of \citeauthor{bergum2025unexpectedinformative} consists of EEG recordings collected from 24 participants, each completing 72 code comprehension trials, resulting in an initial total of 1\,728 trials~\cite{bergum2025unexpectedinformative}. After preprocessing, artifact removal, and synchronization with eye-tracking data performed by \citeauthor{bergum2025unexpectedinformative}, their dataset provides 1\,432 high-quality samples usable as foundation for our analysis of RQ\textsubscript{2} and RQ\textsubscript{3}. Each of these 1\,432 trials was annotated as either \emph{confusing} or \emph{clean} based on whether the code snippet contained an atom of confusion.

\subsubsection{EEG Preprocessing}

To accurately retrieve EEG signals time-locked to fixations on confusing code regions, we first performed the epoching of the raw EEG data into time-locked segments around fixation events. For this, we used a slight modification of the algorithm used by \citeauthor{bergum2025unexpectedinformative} for fixation selection. Our algorithm identifies the first fixation that is closest to the area of interest (and not just the first fixation within the bounding box around this defined area), which more accurately corresponds to the LLM receiving this token as input. The resulting epochs contained data from 27 EEG electrodes, spanning from -300 to 1\,000\,ms at a sampling rate of 500\,Hz, and we used the mean EEG signal for -300--0\,ms as baseline. Building on this fixation selection, we refined our analysis by focusing specifically on the frontal midline electrode (Fz) within a 400--800\,ms window post-fixation, where the late frontal positivity identified by \citeauthor{bergum2025unexpectedinformative} was most pronounced~\cite{bergum2025unexpectedinformative}. For each atom, we followed the standard procedure of averaging the corresponding EEG epochs across participants to enhance the signal-to-noise ratio. We then extracted the mean amplitude for statistical comparison, as established for ERP analysis~\cite{luck2014erp9}.

\subsection{LLM-Based Perplexity Analysis}

\subsubsection{Language Model Selection}

We based the selection of language models for our experiments on their performance on Java tasks from the MultiPL-E benchmark\footnote{\url{https://huggingface.co/datasets/nuprl/MultiPL-E/viewer/humaneval-java}}, a multilingual extension of the HumanEval benchmark designed to evaluate the functional correctness of synthesized programs from docstrings. Specifically, we relied on the Big Code Models Leaderboard\footnote{\url{https://huggingface.co/spaces/bigcode/bigcode-models-leaderboard}}, which reports the average \texttt{pass@1} scores for each model on a set of programming tasks across multiple languages. This metric reflects the proportion of generated code completions that are functionally correct on the first attempt. Because our EEG data stems from a study using Java code, we prioritized models with strong Java performance for perplexity evaluation. Accordingly, we selected \texttt{Qwen2.5-Coder-32B}\footnote{\url{https://huggingface.co/Qwen/Qwen2.5-Coder-32B},~\cite{hui2024qwen2}}, which, at the time of writing, held the highest Java \texttt{pass@1} score on the leaderboard at $65.49\%$. This choice ensured that our perplexity evaluations were based on a state-of-the-art model proficient in Java, thereby enhancing the validity of comparisons with neurophysiological responses to source code and improving the relevance of our findings to real-world program comprehension tasks.

\subsubsection{Token-Level Perplexity Calculation}

To quantify model confusion at a fine-grained level, we compute per-token perplexity using the \texttt{Qwen2.5-Coder-32B} model. This approach follows prior work that uses token-level perplexity to analyze code~\cite{xu2024investigatingefficacyperplexitydetecting}. Our procedure for each code snippet is as follows:

\begin{enumerate}[leftmargin=0.5cm]
    \item Parsing and Tokenization. The snippet is first preprocessed to extract predefined atoms of confusion (based on \citeauthor{bergum2025unexpectedinformative}~\cite{bergum2025unexpectedinformative}) and is then tokenized using the model's tokenizer.
    
    \item Token Probability Calculation. For each token position \( i \), the model receives the previous sequence of tokens as context and predicts the probability distribution over the next token. Logits from the model’s output are converted to probabilities via a softmax function, and the probability assigned to the actual next token \( t_i \) is extracted.

    \item Per-Token Perplexity Computation. We then compute the per-token perplexity as the inverse of the predicted probability for all tokens in the snippet, producing a sequence of perplexity values.
\end{enumerate}

\subsubsection{Calculation of Perplexity for Snippets and Multiple Tokens}
\label{sec:snippet_level_vs_aoi_calculation}

In addition to token-level analysis, we aggregate the perplexity values of multiple tokens to capture model confusion over the code snippets and over areas of interest  (AOIs, i.e., manually defined atoms of confusion or data-driven regions of confusion). Using the standard sequence-level perplexity definition introduced in Section~\ref{sec:perplexity_background}, we compute the following:

\begin{itemize}[leftmargin=0.35cm]
    \item Average perplexity: the standard sequence-level perplexity of the snippet or region, computed via the inverse geometric mean of token probabilities.
    \item Maximum perplexity: the highest perplexity observed for any token within the snippet or region.
\end{itemize}

We calculated these metrics separately for each code snippet and area of interest. They represent the language model's confusion and are used in subsequent analyses to evaluate whether regions associated with higher model confusion also correspond to higher correlates of human confusion (as indicated by EEG fixation-related potentials).

\subsubsection{Statistical Analysis}

For RQ\textsubscript{1}, we first tested for normality with the Shapiro–Wilk test. Since the differences did not follow a normal distribution, we opted for the non-parametric Wilcoxon signed-rank test. We report $p$-values and the rank-based effect size $r = Z/\sqrt{N}$, where $Z$ is the standardized test statistic and $N$ is the number of paired observations. For RQ\textsubscript{2}, we examined the relationship between model perplexity and neurophysiological correlates of confusion by applying a non-parametric, rank-based correlation metric, specifically Spearman's rank correlation coefficient ($\rho$). For RQ\textsubscript{3}, we applied clustered bootstrapping for Spearman's correlation (10\,000 replicates), resampling at the snippet level to control for non-independence of multiple detected regions of confusion per snippet. Our choice of 10\,000 bootstrap replicates was guided by best practices, which recommend using thousands of replicates to ensure stable confidence intervals and reduce Monte Carlo error~\cite{EfronTibshirani1998Review, Davidson2000, Hesterberg2015}. While prior simulation studies on Spearman's correlation have used 2\,000 bootstrap replicates when feasible~\cite{Ruscio2008}, larger values offer improved precision and are increasingly adopted~\cite{Esarey2019}. We report $95\%$ confidence intervals to support robust statistical inference~\cite{francis2018clusteredbootstrapping}.

\subsection{Data-Driven Detection of Regions of Confusion}
\label{sec:data_driven_rocs}

For RQ\textsubscript{3}, we explore whether language model perplexity can serve as a predictive signal for human confusion. To this end, we developed a data-driven method for identifying regions of confusion in code. Our method detects areas of locally-high perplexity and refines them through lexical, syntactic, and structural post-processing. The procedure consists of four main steps.

\subsubsection{Peak Detection} 

Given a tokenized code snippet with token-level perplexity values, we identify local maxima using the promi-nence-based peak detection algorithm from \texttt{scipy}\footnote{version 1.14.1, ~\cite{2020SciPy}}. Prominence measures how much a peak stands out relative to its surroundings, ensuring that only meaningful perplexity peaks are detected as candidates. Through our testing, we set the prominence to 0.8, which reliably isolates sharp perplexity spikes for our dataset. In addition, we set the detection span to 1, such that each detected peak corresponds to a single high-perplexity token, which then forms the core of a candidate region of confusion.
    
\subsubsection{Lexical and Syntactic Expansion}

We first classify tokens using simple lexical rules to identify their categories. Based on this classification, we then expand each peak region syntactically by merging related tokens into meaningful units. This includes combining variable names with numeric suffixes (e.g., \texttt{V1}), or resolving operators with operands (e.g., \texttt{V1-{}-}).
    
\subsubsection{Merging Regions of Confusion}

We merge adjacent or overlapping perplexity-based spans to form unified regions of confusion based on span proximity and overlap. This prevents fragmentation and captures compound expressions that may collectively contribute to confusion.

\subsubsection{Categorization and Filtering}
\label{sec:roc_categorization}

To better understand the nature of the regions flagged by our data-driven detection pipeline, we automatically annotate each identified merged region of confusion using a parser to identify the syntactic category (e.g., \textit{Identifier}, \textit{Punctuation}) of the covered code fragment. For each region of confusion, we extracted the highest node in the abstract syntax tree (AST) fully spanning the region (but not more) and assigned it a label corresponding to the syntactic construct. We then grouped them into broader categories to facilitate interpretability. The mapping is based on typical program comprehension constructs. For example, both \texttt{if} and \texttt{for} constructs fall under \textit{Control Flow}, while operators such as \texttt{++} and \texttt{==} are categorized as \textit{Operator}.

We applied this categorization to all detected regions of confusion across the dataset. We filtered out the regions of confusion categorized as one of the following categories:
\textit{Type} and \textit{Identifier} regions involved identifiers and type declarations that often appeared early in the snippet where the model lacks context, leading to artificially high perplexity that does not necessarily indicate confusion.
Constructs of the category \textit{Control Flow} typically span broad regions (e.g., entire conditional blocks), making it difficult to localize the specific source of model confusion.
Constructs of the category \texttt{Punctuation}, as semicolons, were excluded as the small size of the region likely leads to rare fixation.
After filtering, the majority of retained regions of confusion fell into the \emph{Literal}, \emph{Program Structure}, \emph{Expression}, and \emph{Operator} categories. We provide all analyzed snippets including their and average and maximum token-level and snippet-level perplexity values, detected regions of confusion, and included and excluded syntactic categories as well as the entire script to reproduce this step in our replication package.

%% file: results.tex
\section{Results}
\label{sec:results}

In this section, we present the results regarding our research questions. We first compare perplexity between confusing and clean code snippets (RQ\textsubscript{1}), then analyze correlations with EEG signals (RQ\textsubscript{2}), and finally evaluate our data-driven detection of regions of confusion (RQ\textsubscript{3}).

\subsection{Atoms of Confusion \& LLM Perplexity (RQ\textsubscript{1})}

To address RQ\textsubscript{1}, we analyzed whether atoms of confusion result in higher language model perplexity. For illustration, we begin with a qualitative comparison of a clean–confusing snippet pair as illustrated in Figure~\ref{fig:intro_figure}. The token-level perplexity values show a pronounced spike in the regions of the atoms of confusion compared to their clean counterpart, with both average and maximum perplexity increasing substantially (average: $1$ → $9$; maximum: $1$ → $555$). At the snippet level, we observe an increase in average perplexity ($20$ → $26$), while the maximum ($2\,661$) remains unchanged. These pronounced differences suggest that the presence of atoms of confusion contributes substantially to the model's perplexity.

\begin{figure}[t]
    \centering
    \includegraphics[width=\linewidth]{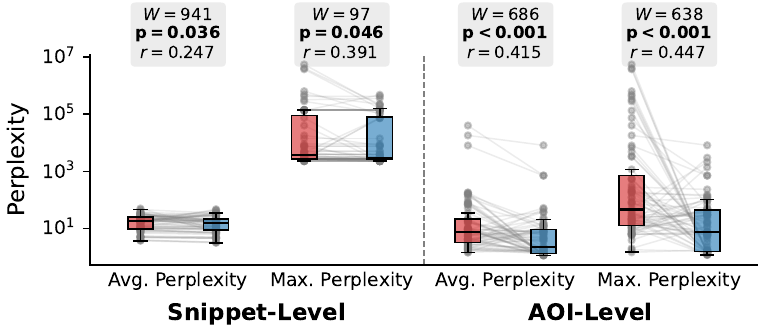}
    \caption{Comparing snippet-level and AOI-level perplexity metrics between \confusing and \clean variants with Wilcoxon signed-rank tests (all statistically significant).}
    \label{fig:ttest_perplexity_metrics}
\end{figure}

To evaluate whether this pattern generally holds, we systematically compared perplexity values between clean and confusing variants across all 72 snippet pairs using a Wilcoxon signed-rank test. As shown in Figure~\ref{fig:ttest_perplexity_metrics}, both snippet-level metrics were statistically significantly higher in confusing snippets (avg.: $W = 941$, $p = 0.036$, $r = 0.25$; max.: $W = 97$, $p = 0.046$, $r = 0.39$).

For a finer-grained analysis, we narrowed our focus to the predefined atoms of confusion and their clean counterparts and evaluated them using the same perplexity metrics, but now at the AOI level (only the tokens related to the atoms and their counterparts). Both average and maximum perplexity were significantly higher in confusing snippets, with stronger effect sizes (avg.: $W=686$, $p<0.001$, $r=0.42$; max.: $W=638$, $p<0.001$, $r=0.45$). The difference in effect size between average and maximum perplexity at the AOI level is relatively small, likely due to the short AOI spans (typically 2–3 tokens) limiting the smoothing effect of averaging. Likewise, the small difference in maximum perplexity between snippet-level and AOI-level is likely because the highest-perplexity token in the snippet often falls within the AOI. Still, maximum AOI-level perplexity offers the most targeted signal and achieves the highest effect size across all metrics making it the preferred measure for the remainder of our analyses.

\RQAnswer{RQ\textsubscript{1}}{We conclude that maximum AOI perplexity provides the most targeted signal of the model's response to known confusing constructs. Consequently, we use maximum AOI perplexity in the remainder of our analyses.}

\subsection{LLM Perplexity \& Neurophysiological Responses (RQ\textsubscript{2})}

To address RQ\textsubscript{2}, we analyzed the relationship between LLM perplexity and human neurophysiological responses in predefined AOIs in terms of fixation-related potentials, as established by \citeauthor{bergum2025unexpectedinformative}~\cite{bergum2025unexpectedinformative}. Each AOI corresponds to an atom of confusion in the confusing variant, or the corresponding region in the clean variant. 

For each atom of confusion, we computed the maximum AOI-level perplexity and identified the longest first fixation inside the region to extract the corresponding EEG signal segment. Based on the preprocessed EEG data of \citeauthor{bergum2025unexpectedinformative}, we computed the mean potential at the Fz electrode in the $400$--$800$\,ms post-fixation window, specifically capturing the previously identified late frontal positivity. 
Next, we computed a Spearman correlation between maximum AOI-level perplexity and mean FRP amplitude. Because the clean and confusing snippets are inherently different, we computed the correlations separately for the clean and confusing variants. 

\begin{figure}[t]
    \centering
    \includegraphics[width=\linewidth]{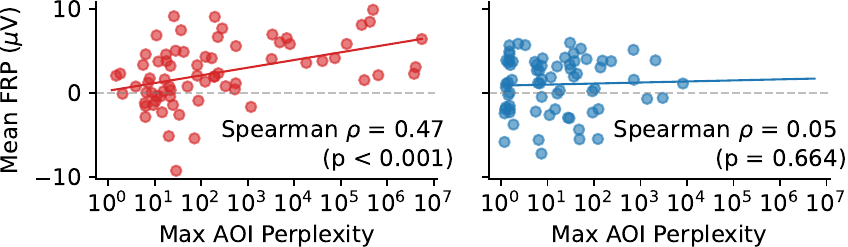}
    \caption{Correlation between maximum AOI-level LLM perplexity and normalized mean FRP amplitude, separately for \confusing (left) and \clean (right) code variants.}
    \label{fig:frp_vs_ppl_scatter}
\end{figure}

Overall, we found positive correlations with varying significance.  For the confusing snippets, we observed a statistically significant moderate positive correlation (Spearman’s $\rho = 0.47$, $p < 0.001$). For the clean snippets, we observed a negligibly positive correlation, which is not statistically significant (Spearman’s $\rho = 0.05$, $p = 0.664$). We visualize these correlations in~\autoref{fig:frp_vs_ppl_scatter}.

\RQAnswer{RQ\textsubscript{2}}{For confusing code, the LLM perplexity significantly correlates with the neurophysiological response of human programmers associated with confusion.}

\subsection{Detecting Confusing Code (RQ\textsubscript{3})}

To evaluate whether LLM perplexity can be used to detect confusing code regions, we implemented a data-driven method to identify such regions directly from token-level perplexity values. Specifically, we located local maxima in perplexity and aggregated them into syntactically coherent spans, which we refer to as \emph{regions of confusion}. Depending on how many perplexity peaks occur in a snippet, the algorithm may identify multiple regions of confusion.

To validate this approach, we compared the perplexity-based regions of confusion to the annotated atoms of confusion introduced by \cite{bergum2025unexpectedinformative}. A detected region was considered to overlap with an atom of confusion if it shared, at least, one token with a known confusing construct in the same code snippet. 

\begin{figure}[h]
    \centering
    \includegraphics[width=0.85\linewidth]{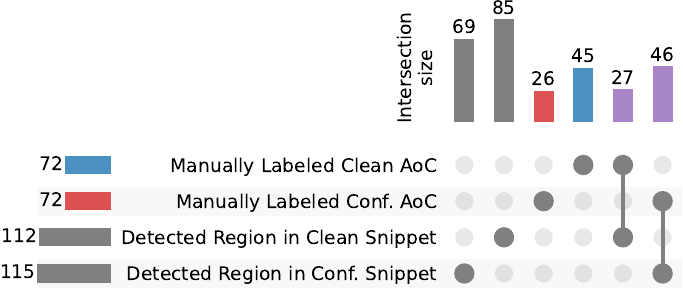}
    \caption{
        UpSet plot showing the overlap between manually annotated atoms of confusion in \clean and \confusing snippets, and automatically detected regions (gray). 
        Horizontal bars indicate the total number per category, while vertical bars represent intersections. Purple bars indicate overlaps.
    }
    \label{fig:upset_overlap}
\end{figure}

The data-driven approach was able to detect regions of confusion in 64\% of the manually labeled atoms of confusion, as well as regions of confusion that are possibly beyond the known atoms of confusion. 
Specifically, this approach detected a total of 227 distinct regions of confusion across the dataset: 115 regions in confusing snippets and 112 in clean snippets, many of which were novel and showed no overlap with manually labeled atoms of confusion. We visualize the overlap between our data-driven regions and the manually annotated atoms of confusion using an UpSet plot in Figure~\ref{fig:upset_overlap}. Notably, the method detected 85 novel regions in confusing snippets and 69 in clean snippets, revealing potential regions of confusion not captured by manual annotation. The method largely overlapped with the manually labeled atoms of confusion in confusing snippets, indicating that a data-driven approach is well able to detect regions that actually contain confusion. However, 26 atoms of confusion were not detected, indicating that some of the confusing constructs could not be captured by our approach. Similarly, 45 out of 72 manually labeled clean atoms were not detected by the method, but 27 detected regions in clean snippets overlapped with manually labeled confusing atoms. These 27 overlaps could be considered false positives in a strict sense.

%This plot shows the total number of items in each category (horizontal bars) and the size of each intersection (vertical bars with connected dots), allowing for a more detailed view of how detected regions align with clean and confusing annotations.

\begin{figure}[t]
    \centering
    \includegraphics[width=\linewidth]{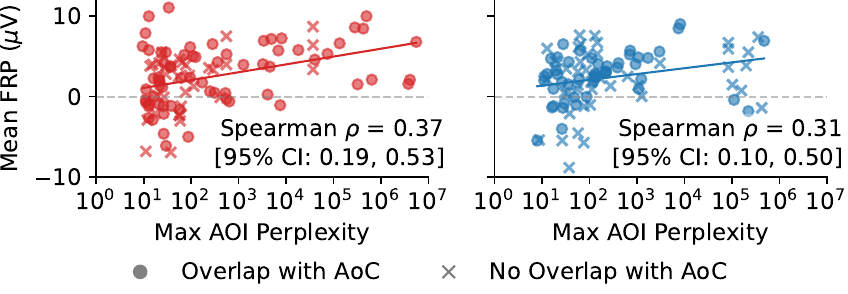}
    \caption{Correlation between maximum AOI-level LLM perplexity and normalized mean FRP amplitude in automatically detected regions of confusion, for \confusing (left) and \clean (right) code variants. Markers distinguish overlap (o) with known atoms vs.\ non-overlap (×).}
    \label{fig:clean_vs_confusing_overlap_vs_nooverlap}
\end{figure}

To better understand whether these newly detected perplexity-based regions of confusion are confusing to humans, we conducted a similar analysis to the one employed to answer RQ\textsubscript{2}. Specifically, we extracted for all detected regions the closest fixation and computed the mean EEG amplitude in the $400$--$800$\,ms post-fixation window at the Fz electrode. 
Our clustered bootstrap Spearman analysis (resampling by snippet to account for multiple regions per snippet) shows a robust correlation with neurophysiological correlates of confusion in the confusing snippets ($\rho = 0.37,\ 95\% \ \text{CI} = [0.19, 0.53])$. Note that, even in clean snippets, the data-driven approach detected regions align with FRP responses ($\rho = 0.31,\ 95\% \ \text{CI} = [0.10, 0.50]$).
We visualize the correlations of the perplexity in regions of confusion with the human response in~Figure~\ref{fig:clean_vs_confusing_overlap_vs_nooverlap}.

To understand whether the novel detected regions or manually labeled regions drive the robust correlation, we computed correlations separately for regions that did and did not overlap with previously annotated atoms of confusion. Overall, there are positive trends across all subgroups, but the stronger correlations appear in overlapping regions, both in clean and confusing snippets. Table~\ref{tab:spearman_results} reports each Spearman rank correlation including 95\% confidence intervals for all four subgroups.

\RQAnswer{RQ\textsubscript{3}}{The data-driven, perplexity-based method effectively detects confusing code regions, discovering nearly two-thirds of known confusing constructs in confusing variants. The detected regions of confusion show positive correlations with human confusion, even in (based on the literature) supposedly clean code snippets.}

\begin{table}[t]
\centering
\caption{Spearman rank correlations ($\rho$) between maximum AOI Perplexity and mean FRP response, with 95\% confidence intervals computed via clustered bootstrapping (10\,000 resamples) for sub-groups. * indicates significance (0 $\notin \text{CI}$).}
\label{tab:spearman_results}
\small
\begin{tabular}{@{}lcl@{}}
\toprule
\textbf{Group} & \textbf{Spearman $\rho$} & \textbf{95\% CI}  \\ \midrule
Overlap with \clean AoC & 0.38 & \textbf{[0.10, 0.70]*} \\
No overlap with \clean AoC & 0.24 & [-0.05, 0.49] \\
Overlap with \confusing AoC & 0.33 & \textbf{[0.12, 0.52]*} \\
No Overlap with \confusing AoC & 0.36 & [-0.05, 0.71] \\ \bottomrule
\end{tabular}
\end{table}

%% file: discussion.tex
\section{Discussion}
\label{sec:discussion}
In this section, we discuss our results and their implications by research question.

\subsection{Atoms of Confusion \& LLM Perplexity (RQ\textsubscript{1})}

Our analysis reveals significant differences in perplexity measures between clean and confusing code variants at both snippet and AOI levels. However, maximum AOI-level preplexity provides the strongest and most robust effect ($r = 0.447$), likely because it is not influenced by noise within or outside the atom of confusion. As snippet length increases---such as when analyzing full functions or files---we expect both average and maximum perplexity at the snippet level to lose their effectiveness as reliable indicators of confusion. Likewise, we expect average AOI-level to lose its effectiveness for confusing code constructs spanning more tokens.
%the effect size for snippet-level average perplexity is relatively small, suggesting that the signal is primarily driven by localized constructs within otherwise similar code.
%While snippet-level maximum perplexity showed a somewhat stronger effect, it remains below the AOI-level values and is less robust overall. In longer code spans, snippet-level maximum perplexity becomes increasingly sensitive to outliers, making it prone to noise and unreliable as a confusion indicator. As snippet length increases---such as when analyzing full functions or files---we expect both average and maximum perplexity at the snippet level to lose their effectiveness as reliable indicators of confusion.
%In contrast, focusing on predefined AOIs corresponding to atoms of confusion revealed stronger and more consistent effects. Both average and maximum perplexity were significantly higher in confusing code, with maximum AOI perplexity yielding the highest effect size across all metrics ($r = 0.447$). These findings indicate that LLM confusion is most effectively captured through localized spikes in perplexity, and that the maximum perplexity within a targeted region serves as a particularly sensitive indicator. 
Visual inspection of token-level perplexity distributions further corroborates these quantitative findings, showing clear alignment between perplexity peaks and the location of known confusing code constructs (see~\autoref{fig:intro_figure}(b)).

\paragraph{Implications for Research}
These results support the interpretation that LLM confusion, when triggered by atoms of confusion, is a localized phenomenon rather than a characteristic that spans entire snippets. Prioritizing such fine-grained, token-level perplexity metrics enables more accurate modeling and alignment between LLM and human confusion. Future work shall explore the variability of LLM perplexity responses across different categories of confusing constructs and investigate cases where perplexity peaks fall outside annotated AOIs, potentially reflecting deeper model confusion, ambiguous semantics, or gaps in existing atoms of confusion datasets. Additionally, alignment between perplexity peaks and annotated atoms of confusion may be further improved through prompting strategies that guide the model's attention or through supervised fine-tuning on annotated atoms of confusion data.

\paragraph{Implications for Practitioners}
From a practical standpoint, maximum AOI perplexity provides an interpretable and computationally efficient signal for identifying regions of potential confusion in code. Since perplexity can be obtained directly from pretrained language models without requiring additional instrumentation or manual labeling, it offers a scalable and lightweight approach for real-time confusion detection. Such signals have promising applications in developer tools and educational environments, where highlighting cognitively demanding code regions can assist in code review, debugging, and learning processes.

\subsection{LLM Perplexity \& Neurophysiological Responses (RQ\textsubscript{2})}

Our results demonstrate a strong and statistically significant positive correlation between the maximum perplexity predicted by the language model and the amplitude of fixation-related potentials in response to confusing code variants (Spearman’s $\rho = 0.47$, $p < 0.001$). In contrast, clean code variants showed only a negligibly positive correlation ($\rho = 0.05$, $p = 0.664$). This provides compelling evidence that LLM confusion is meaningfully aligned with human confusion during program comprehension, bridging computational measures and neurophysiological responses.

To the best of our knowledge, this is the first direct empirical validation linking LLM confusion and direct, neurophysiological correlates of human confusion in the domain of program comprehension. Thus, the results have far-reaching implications for both research and practice:

\paragraph{Implications for Research}
From a research perspective, these findings contribute to an emerging line of research advocating for the use of neurophysiological measures to validate and interpret LLM behavior in software engineering contexts~\cite{srikant2022convergent}. They highlight the prospects of incorporating neurophysiological signals into LLM evaluation and fine-tuning frameworks, potentially advancing model interpretability and alignment with human neurophysiological responses~\cite{toneva2019interpreting,aw2023training,oota2024speech}. 
Furthermore, our results show that, when modeling human programmers, LLMs can serve as a viable surrogate model and a foundation for deriving and verifying hypotheses about programmer behavior when designing empirical experiments. In the future, this could even be extended by providing the approach with more information about the context, such as programmer experience (e.g., familiarity with certain constructs) or reading behavior (e.g., previously regarded tokens).

\paragraph{Implications for Practitioners}
For practitioners, these findings provide evidence for the observation that LLMs are well able to collaborate with humans by being sensitive to human confusion. Additionally, they position perplexity as a reliable, lightweight measure for cognitive difficulty. Compared to behavioral metrics such as reading times or error rates, perplexity offers a scalable, token-level, and model-based signal that can be computed efficiently on large code bases.
This opens avenues for developing human-aligned programming tools---such as adaptive integrated development environments and educational systems---that dynamically respond to perceived confusion, enhancing usability and learning efficiency.

\subsection{Detecting Confusing Code (RQ\textsubscript{3})}

To move beyond the limitations of manually annotated atoms of confusion, we devised a fully automatic method that identifies potentially confusing regions in source code by locating local maxima in token-level LLM perplexity. The method uncovered the majority of manually-labeled confusing snippets, while also detecting additional regions of confusion in supposedly clean snippets (cf.~\autoref{fig:upset_overlap}). It is important to note that the clean snippets are not guaranteed to be completely ``free of confusion'', not even in the annotated area. Instead, they only proclaim to be less confusing than the corresponding atom of confusion. For example, bitwise operations, such as \texttt{0b0010 \& 0b0100} (\textcolor{clean}{clean}) versus \texttt{12 \& 3} (\textcolor{confusing}{confusing}), may both plausibly generate confusion, explaining the model's perplexity peaks in clean snippets. Thus, overlaps may reflect subtle or context-dependent confusion cues rather than outright errors in the automatic detection. 

%From a methodological perspective, we used clustered bootstrapping by snippet to control for repeated measures, and we reported 95\% confidence intervals to ensure robust inference. This supports the reliability of the observed moderate correlations, which are notable given the inherent variability of EEG data and the size of the snippets.

To assess the cognitive relevance of these data-driven regions of confusion, we aligned them with fixation-related potentials from EEG recordings. We found no significant correlations for clean snippets when relying solely on manually annotated clean atom of confusion counterparts in RQ\textsubscript{2}. However, with the inclusion of the 85 additional regions detected by our method in clean snippets---alongside some overlapping snippets with annotated atoms---we now observe a moderate and significant correlation, suggesting that clean snippets can still contain subtle confusing constructs. This shift suggests that while the novel, non-overlapping regions alone do not yield a significant correlation (cf.\ Table~\ref{tab:spearman_results}), their combination with overlapping regions, especially after removing many from the original clean set, produces a more neurophysiologically grounded predictive signal. 

%Overall, the findings suggest that perplexity-based detection captures a meaningful subset of known confusion points. At the same time, it also reveals novel, unannotated regions and subtle ambiguities, reinforcing its potential to augment and refine established regions of confusion as well as to detect context-dependent confusion.

Overall, the findings suggest that the effectiveness of our perple-xity-based method lies not in novel detections alone, but in its ability to identify a targeted subset of regions in a data-driven way---some overlapping with atoms of confusion, others newly discovered---while excluding less informative ones. This selective refinement reveals confusing regions that align with neurophysiological signals and may be missed by manual annotation, further supporting perplexity as a meaningful and grounded predictor of confusion in source code, particularly when used to guide region selection in tandem with, rather than independently from, existing curated sets of atoms of confusion.

\paragraph{Implications for Research}
Together with the results from RQ\textsubscript{2}, we conclude that the relationship between perplexity and fixation-related potentials suggests that confusion is better conceptualized as a continuous spectrum, rather than a fixed set of discrete and equivalent constructs. These findings reinforce the idea that our perplexity-based approach captures varying levels of neurophysiological responses, enabling scalable and empirically validated detection of nuanced confusing regions. 
On a broader scale, this work provides a research avenue for a more human-aligned LLM working in concert with programmers. Future work shall investigate how LLMs might reflect other neurophysiological responses in their internal state. One example might be semantic implausibility, which elicits N400 responses in humans when reading natural language or code~\cite{kuperberg2020twoposn400,kuo2024violationserp}, and is sometimes also associated with increased perplexity~\cite{wilcox2020predictive,salicchi2025not,lopes2024language}. Incorporating automatically detected regions of confusion into research workflows---such as those used to study atoms of confusion or other confusing constructs---could support the development of hybrid frameworks that reduce reliance on manual curation through corpus studies and surveys.

\paragraph{Implications for Practitioners}
This opens promising avenues for support tools that adapt dynamically to the programmer's cognitive state by allowing data-driven feedback methods to flag or revise cognitively demanding code based on LLM perplexity. It also facilitates the development of tools that automatically refactor confusing code or highlight areas where programmers may need additional guidance.
Future work shall refine detection thresholds to optimize sensitivity and specificity of the detection algorithm by varying prominence (e.g., to filter out regions with little to no confusion). Additionally, it might require deeper investigation of how larger context (i.e., bigger code snippets) influences the effect of atoms of confusion or creates further confusion regions in human programmers, whether LLMs also reflect these confusion patterns in their perplexity, and whether they might provide assistance in their detection. These findings should also be validated for a large and diverse population of programmers varying in experience and knowledge, and whether prior knowledge can be inserted into the model context to suit individual needs of programmers.

%% file: threats.tex
\section{Threats to Validity}
\label{sec:threats}

Our study combined the elements of confusing code, human thinking, LLMs, and their linkage, which all come with their own sets of threats to validity. Here, we discuss how we mitigated these threats.

\subsection{Construct Validity}

Our study relies on the constructs of ``comprehending`` code, which is ``confusing`` to the LLM and humans. Both are theoretical constructs that are inherently difficult to measure. Therefore, we relied on well-established operationalizations to minimize threats to construct validity. Specifically, we used the validated set of atoms of confusion~\cite{gopstein2017atoms,langhout2021atoms,bergum2025unexpectedinformative} as trigger for confusion. The used measure of human confusion, the fixation-related potential approach, is also a well-established indicator of cognitive processes in psychology and related fields~\cite{hutzler2007welcome,degno2020frpinreading}.

Similarly, we used perplexity as a commonly used, fast and scalable measure of uncertainty for the LLM~\cite{shorinwa2025surveyuncertainty,xiong2024efficient}, which we believe is the best comparable metric for LLM confusion. For our RQ\textsubscript{3}, we detected regions of confusion based on peaks in the LLM perplexity and expanded it by the underlying code construct. This approach requires setting parameters for the peak detection and thresholds on how much to expand to the surrounding tokens. We minimized this threat to construct validity by using reasonable values and providing the full analysis script in the replication package.

Finally, we hypothesize in this paper that perplexity for LLMs and FRP-measured late frontal positivity for humans are related constructs. The data support this hypothesis, but further studies shall corroborate it.

\subsection{Internal Validity}

We view the diversity of the set of code snippets as the main threat to internal validity. Atoms of confusion cover several different constructs. We rely on the EEG data of~\citeauthor{bergum2025unexpectedinformative}~\cite{bergum2025unexpectedinformative}, which treat all snippets as equally confusing. However, as shown in~\autoref{fig:upset_overlap}, there is a variety in the constructs and how well we can detect them with perplexity. It is possible that the different constructs, or other confounding factors, affect the found link between human confusion and LLM confusion.

An additional threat to internal validity is the size of the dataset of~\citeauthor{bergum2025unexpectedinformative}~\cite{bergum2025unexpectedinformative}, on which we relied for our study. While providing a substantial contribution with their dataset and measuring the confusion across all atoms (with over 600 epochs per condition), the small sample size up to 12 epochs for a single atom is challenging for our analysis. We minimized this threat by using only the data from electrode Fz (which showed the most prominent signal) and by averaging the EEG signal across the time period of 400--800\,ms instead of performing a more detailed analysis over this time span.

\subsection{External Validity}

On the LLM side, we focused on \texttt{Qwen2.5-Coder-32B} due to its performance on Java coding tasks. To minimize the potential that our findings are specific to this LLM model, we ran our analysis with several other top-ranking models of varying sizes (i.e., \texttt{Deepseek-Coder~1.3B}, \texttt{Llama~3.2~3B}, \texttt{Artigenz-Coder-DS~6.7B}, \texttt{Deepseek-Coder~6.7B}, \texttt{Yi-Coder~9B}), which showed very similar results. We include those results in our replication package. 

Furthermore, natural language studies have shown that there is a turning point, at which larger models become less aligned with human natural language processing~\cite{oh2023largerpoorer,lin2025largerfMRI}. Thus, a smaller coding model may show a closer link with how humans comprehend code. However, finding this turning point is beyond the scope of our study.

On the human side, we refer to the threats to validity of the used dataset of~\citeauthor{bergum2025unexpectedinformative}~\cite{bergum2025unexpectedinformative}. Most critically, it is possible that the found effects of confusion via FRP are specific for the participant sample of intermediate programmers and may not generalize to programmers of all experience levels or backgrounds.

Finally, we used atoms of confusion as the best way to obtain a strong signal. It is possible our findings are specific to the confusion triggered by atoms of confusion, and may not generalize to all confusing code constructs. Further studies shall investigate whether our results showing a link between LLM and human thinking holds up in other code, tasks, and contexts.

%% file: conclusion.tex
\section{Conclusion}
\label{sec:conclusion}

As humans increasingly collaborate with LLM-based programming assistants, it is critical to investigate whether and to what extent LLMs and humans align in this task. To this end, we conducted an empirical study examining an
LLM comprehending clean and confusing code (in terms of LLM perplexity) as compared to human programmers comprehending the same code (in terms of neurophysiological responses). We found that LLM perplexity spikes correlate both in terms of location and amplitude with human neurophysiological responses that indicate confusion.
This suggests that LLMs and humans are similarly confused about the same code. Thus, we have devised a data-driven, LLM-based approach to identify code regions of confusion that are associated with human confusion.

Our study revealed an alignment between LLM perplexity and human neurophysiological responses, which represents an important step toward establishing LLMs as surrogate models of human program comprehension. This opens new avenues for exploring the underlying cognitive processes of human program comprehension and supports the development of LLMs to be more closely aligned with human comprehension patterns, such as other sources of confusion. We lay the groundwork for data-driven feedback methods that, using LLM perplexity, detect and revise cognitively demanding code. This approach can be optimized to generalize for larger code snippets and individual programmers, which facilitates the development of tools that automatically refactor confusing code or highlight areas in which programmers may need additional guidance.

%% file: acknowledgments.tex
\begin{acks}

% TODO when accepted, also ask Mariya for her grants
This work has been supported by the European Union as part of ERC Advanced Grant ``Brains On Code'' (101052182).
\end{acks}